# Surface Flux Transport Modeling Using Physics-informed Neural Networks

Jithu J Athalathil[1], Bhargav Vaidya[1], Sayan Kundu[1,2], Vishal Upendran[3,4], and Mark C. M. Cheung[5]
[1] Department of Astronomy Astrophysics and Space Engineering, Indian Institute of Technology Indore, Khandwa Road, Simrol, Indore 453552, India
[2] Department of Physics, University of Bath, Claverton Down, BA2 7AY, UK
[3] Bay Area Environmental Research Institute, Moffet Field, CA, USA
[4] Lockheed Martin Solar and Astrophysics Laboratory, Palo Alto, CA, USA
[5] CSIRO, Space & Astronomy, PO Box 76, Epping, NSW 1710, Australia

## Abstract

Studying the magnetic field properties on the solar surface is crucial for understanding the solar and heliospheric activities, which in turn shape space weather in the solar system. Surface flux transport (SFT) modeling helps us to simulate and analyze the transport and evolution of magnetic flux on the solar surface, providing valuable insights into the mechanisms responsible for solar activity. In this work, we demonstrate the use of machine learning techniques in solving magnetic flux transport, making it accurate. We have developed a novel physics-informed neural network (PINN)-based model to study the evolution of bipolar magnetic regions using SFT in one-dimensional azimuthally averaged and also in two dimensions. We demonstrate the efficiency and computational feasibility of our PINN-based model by comparing its performance and accuracy with that of a numerical model implemented using the Runge–Kutta implicit–explicit scheme. The mesh-independent PINN method can be used to reproduce the observed polar magnetic field with better flux conservation. This advancement is important for accurately reproducing observed polar magnetic fields, thereby providing insights into the strength of future solar cycles. This work paves the way for more efficient and accurate simulations of solar magnetic flux transport and showcases the applicability of PINNs in solving advection–diffusion equations with a particular focus on heliophysics.

*Unified Astronomy Thesaurus concepts:* Solar physics (1476); Solar magnetic fields (1503); Solar surface (1527); Solar active regions (1974)

## 1. Introduction

The dynamic processes on the solar surface exert a significant influence on the magnetic properties of the Sun. It is widely accepted that the generation of magnetic fields in the convection zone is the result of a dynamo mechanism (K. Petrovay 2000; M. Ossendrijver 2003; D. Nandy 2009; A. R. Choudhuri 2011, 2014; P. Charbonneau 2020). Solar active regions (ARs) represent areas on the Sun's surface exhibiting intense magnetic activity. They often manifest as dark patches known as sunspots (S. K. Solanki 1993, 2003; A. Ruzmaikin 2001; N. O. Weiss 2006). These regions form as a result of strong toroidal flux tubes rising through the convection zone and emerging as bipolar magnetic regions (BMRs) on the photosphere. They deposit magnetic flux on the photosphere, which subsequently migrates toward the poles through the meridional flow. This results in the regeneration of the poloidal magnetic field of the Sun. Eventually, differential rotation comes into play and disrupts the poloidal field, giving rise to a toroidal field, and thereby the cycle continues (Y. Fan 2009; R. F. Stein 2012; M. C. M. Cheung et al. 2016; P. Charbonneau 2020). Simulating the surface transport of the magnetic flux provides valuable insights into the mechanisms responsible for solar activity. Surface flux transport (SFT) modeling enables us to simulate and analyze the transport and evolution of magnetic flux on the solar surface (N.R. Sheeley 2005; J. Jiang et al. 2014; A. R. Yeates et al. 2023). Magnetic field modeling aids in constraining the solar dynamo models. Additionally, it also serves as an initial condition for extrapolating the solar magnetic field to the heliosphere (R. H. Cameron et al. 2012; A. Lemerle & P. Charbonneau 2017; G. Hazra 2021). These simulations facilitate forecasting the magnetic activity of the upcoming solar cycle (H. Iijima et al. 2017; L. A. Upton & D. H. Hathaway 2018; P. Bhowmik & D. Nandy 2018).

The SFT model is based on the idea that the solar surface-observed radial magnetic flux behaves as a passive scalar field (R. B. Leighton 1964). The magnetic field observed on the solar surface exhibits only a slight tilt relative to the radial direction (S. K. Solanki 1993; V. Martinez Pillet et al. 1997) and therefore can be safely considered to be oriented radially (Y.-M. Wang et al. 1992). As a result, the flux transfer of the large-scale solar magnetic field on the solar surface can be described using only the induction equation for the radial component of the magnetic field ($B_r$; R. B. Leighton 1964; C. R. DeVore et al. 1984; N. R. Sheeley et al. 1985; D. Orozco Suárez et al. 2007).

Recent advances in SFT simulation methodologies include the advective flux transport (AFT) models, where convective motions are directly integrated into the simulation (L. Upton & D. H. Hathaway 2013). Additionally, the utilization of spherical harmonics, attributed to the spherical geometry, is another notable approach (D. H. Mackay et al. 2002). In this method, the radial component of the magnetic field is included through the superposition of various spherical harmonic components. There have been efforts to use an algebraic method to predict the solar axial dipole moment at solar minimum (M. Nagy et al. 2020; K. Petrovay et al. 2020). Besides empirical relations, various data assimilation techniques have been employed in the literature. One such approach involves utilizing magnetogram data from subsequent time steps to determine magnetic potential, which is then incorporated as an input for simulation (A. R. Yeates et al. 2015). These techniques have found application in numerous studies, including the prediction of solar cycles (L. Upton & D. H. Hathaway 2014; P. Bhowmik & D. Nandy 2018; S. Pal et al. 2023; B. K. Jha & L. A. Upton 2024).

In the past few decades, artificial neural networks have found application in solving problems across a wide spectrum of fields ranging from robotics to healthcare (L. Alzubaidi et al. 2021; I. H. Sarker 2021). About half a decade ago, a significant method emerged for solving partial differential equations (PDEs) using artificial neural networks. (M. Raissi et al. 2019) introduced this technique as physics-informed neural networks (PINNs). This method uniquely combines the strengths of neural networks with the fundamental physical laws governing the systems they model. Consequently, it has been applied across various scientific and technological fields to develop more accurate and robust models for the physical systems in question. In the field of heliophysics, PINNs have been applied to create models for nonlinear force-free fields (NLFFFs; R. Jarolim et al. 2023, 2024), address the inverse problem related to radiation belt electron transport using data from Van Allen probes (E. Camporeale et al. 2022), and investigate magnetic reconnection in coronal fields (H. Baty 2024). The PINN framework has also been employed for various other problems, including advection–diffusion equations (A. T. A. Gomes et al. 2022; S. R. Vadyala et al. 2022). More details can be found in the reviews of PINNs and their versatile applicability (e.g., G. E. Karniadakis et al. 2021; S. Cuomo et al. 2022; H. Baty 2024, and references therein).

SFT simulations involve numerous physical processes characterized by distinct spatial and temporal scales, posing computational challenges and expenses. Machine learning, particularly neural networks can be effectively used to obtain higher accuracy. In this work, we aim to address the scale dependencies using the PINN method. Due to the meshless characteristic, PINNs eliminate the need for traditional grids, thereby removing grid dependencies (M. Raissi et al. 2019). We apply the PINN framework to model solar surface flows and subsequently compared our results with those obtained from a newly developed numerical code.

The article is organized as follows. The SFT equation along with model assumptions are elaborated in Section 1.1. Section 2 provides an in-depth explanation of the numerical methodologies that were employed for conducting our simulations. We have dedicated Section 3 to elucidate the fundamental architecture and optimization techniques applied in the PINN framework. The validation process used to verify the codes is explained in Section 4. Section 5 encapsulates comprehensive descriptions of the solutions using various methods to the 1D and 2D SFT equations including the application to study the evolution of a particular AR. Finally, Section 6 summarizes the work conducted in this study.

### 1.1. Physical Equations

SFT modeling is a technique employed to characterize the movement of radial magnetic flux on the solar surface, driven by various physical processes. The temporal evolution of the radial component of the magnetic field on the solar surface, $B_r(\lambda, \phi, t)$, where $\lambda$, $\phi$, and $t$ represent the latitude, longitude, and time, respectively, accounts for various plasma flows on the solar surface as primary drivers of magnetic flux transport. This includes the meridional flow directed poleward, the differential rotation, and the diffusive cancellation arising from flux merging as indicated in Equation (1). Additionally, the emergence of new ARs over time $t$ is incorporated into the model through the inclusion of a source term denoted as $S(\phi, \lambda, t)$.

$$\frac{\partial B_r}{\partial t} + \frac{1}{R_\odot \cos \lambda} \frac{\partial}{\partial \lambda}(B_r u(\lambda)\cos(\lambda)) + \Omega(\lambda)\frac{\partial B_r}{\partial \phi}$$
$$= \frac{\eta}{R_\odot^2 \cos^2 \lambda} \frac{\partial^2 B_r}{\partial \phi^2} + \frac{\eta}{R_\odot^2 \cos \lambda} \frac{\partial}{\partial \lambda}\left(\cos \lambda \frac{\partial B_r}{\partial \lambda}\right)$$
$$- \frac{B_r}{\tau} + S(\phi, \lambda, t), \quad (1)$$

where $\lambda$ is the latitude ($-90°$ to $+90°$), $\phi$ is the longitude ($0°$–$360°$), and $R_\odot$ represents the solar radius. Here, $u(\lambda)$ represents the advection velocity associated with the meridional circulation and is given by

$$u(\lambda) = \begin{cases} u_0 \sin\left(\frac{\pi \lambda}{\lambda_0}\right) & \text{if } |\lambda| \leqslant \lambda_0, \\ 0 & \text{otherwise} \end{cases} \quad (2)$$

where the value of $u_0 = 12.5 \text{ ms}^{-1}$ and $\lambda_0 = 75°$ is the latitude above which the meridional flow becomes negligible. The differential rotation $\Omega(\lambda)$ is employed by the latitude-dependent profile (H. B. Snodgrass 1983)

$$\Omega(\lambda) = 0.18 - 2.36 \sin^2 \lambda - 1.787 \sin^4 \lambda \text{°/day}. \quad (3)$$

We have chosen diffusion coefficient $\eta = 500 \text{ km}^2 \text{ s}^{-1}$ (C. R. DeVore et al. 1984) and the decay constant $\tau = 5$ yr in our calculations (A. Lemerle et al. 2015; K. Petrovay & M. Talafha 2019). The source (S) of the magnetic field differs depending on the situation and will be described in the context of each specific case. The 1D SFT equation obtained by taking the longitude ($\phi$) average of Equation (1 on both sides takes the form:

$$\frac{\partial B_r}{\partial t} = -\frac{1}{R_\odot \cos \lambda} \frac{\partial}{\partial \lambda}(B_r u(\lambda)\cos(\lambda))$$
$$+ \frac{\eta}{R_\odot^2 \cos \lambda} \frac{\partial^2 (B_r \cos \lambda)}{\partial \lambda^2} - \frac{B_r}{\tau} + S(\lambda, t). \quad (4)$$

The above equation has been studied for predicting the strength and timing of sunspot cycle (I. Baumann et al. 2006; R. H. Cameron et al. 2012; H. Iijima et al. 2017; L. A. Upton & D. H. Hathaway 2018; P. Bhowmik & D. Nandy 2018; G. Hazra 2021; J. Jiang et al. 2022; E. M. Golubeva et al. 2023).

The SFT equation is a linear differential equation and hence the solutions can be expressed as linear superpositions. This implies that if we have solutions for two or more different initial conditions, the solution for the combined initial conditions can be calculated by superposing the individual solutions. This ensures a basis set of solutions that can be superposed to address any complex scenario. This property can be exploited using Green's function approach, which is a powerful method for solving linear differential equations. This interesting connection will be discussed in more detail later.

## 2. Numerical Implementation of SFT

The evolution of the magnetic field with latitude (1D) is realized through the discretization of the azimuthally averaged SFT equation (Equation (4)), employing a finite-volume method. We have used two different numerical schemes, viz the (a) Explicit scheme and (b) Runge–Kutta implicit–explicit (RK-IMEX) scheme, to evolve the magnetic field using the SFT equation. To apply these schemes, we consider the SFT equation as

$$\frac{\partial B_i}{\partial t} = -A_i + R_i, \quad (5)$$

where $A_i$ represents the advection term (first term on the right-hand side of Equation (4)) and $R_i$ represents the diffusion, source, and decay terms summed together (last three terms on the right-hand side of Equation (4)) for the spatial point $\lambda_i$. In the following subsections, the methods adopted to calculate these terms ($A_i$ and $R_i$) for each scheme (Explicit and RK-IMEX) are described.

### 2.1. Scheme 1: Explicit Scheme

The temporal evolution of the magnetic field using the 1D SFT equation (Equation (4)) is achieved using the Runge–Kutta 2 (RK2) time stepping. The RK2 coefficients $K_1$ and $K_2$ for the $i$th grid point are calculated as

$$K_1 = -A_i(t_n, B_i^n) + R_i(t_n, B_i^n) \quad (6)$$

$$K_2 = -A_i(t_n + \Delta t, B_i^n + \Delta t\, K_1) + R_i(t_n + \Delta t, B_i^n + \Delta t\, K_1), \quad (7)$$

where $t_n$ denotes the time of the $n$th time step, $B_i^n$ denotes the magnetic field at the $i$th grid point for the $n$th step and $\Delta t$ being the time step. The temporal update of the magnetic field for the $(n+1)$th time step is calculated as

$$B_i^{n+1} = B_i^n + \Delta t \left( \frac{K_1 + K_2}{2} \right). \quad (8)$$

This method ensures the second-order accuracy in time.

The advection part ($A_i$) is calculated using the upwind scheme with a Van Leer flux limiter (B.V. Leer 1977). The advection update is calculated using the advection flux $\hat{F}(B, u)$ that depends on the advection velocity ($u$; see Equation 2) and magnetic field ($B$) as shown below:

$$A_i = \frac{\hat{F}_{i+1/2} - \hat{F}_{i-1/2}}{\Delta \lambda}. \quad (9)$$

Here, $\Delta \lambda$ denotes the latitude grid spacing. The flux values on the left and right grid walls of the $i$th grid cell are denoted by $\hat{F}_{i+1/2}$ and $\hat{F}_{i-1/2}$, respectively. In order to ensure stability, ($\Delta t$) is selected so as to satisfy the Courant–Friedrichs–Lewy (CFL) condition.

The advection flux follows an upwind selection rule:

$$\hat{F}_{i+\frac{1}{2}} = \begin{cases} u(\lambda_{i+\frac{1}{2}}) B^L_{i+\frac{1}{2}} & \text{if } u(\lambda_{i+\frac{1}{2}}) > 0 \\ u(\lambda_{i+\frac{1}{2}}) B^R_{i+\frac{1}{2}} & \text{if } u(\lambda_{i+\frac{1}{2}}) \leqslant 0 \end{cases}, \quad (10)$$

where $\lambda_{i+\frac{1}{2}} = \lambda_i + \frac{\Delta \lambda}{2}$. The left and the right values of the magnetic field $B^L_{i+\frac{1}{2}}$ and $B^R_{i+\frac{1}{2}}$ are calculated as

$$B^L_{i+\frac{1}{2}} = B_i^n + \frac{\delta B_i}{2}, \quad (11)$$

$$B^R_{i+\frac{1}{2}} = B_{i+1}^n - \frac{\delta B_{i+1}}{2}. \quad (12)$$

Here the harmonic mean slope limiter $\delta B_i$ is given by

$$\delta B_i = \begin{cases} \dfrac{2 \Delta B_{i+\frac{1}{2}} \Delta B_{i-\frac{1}{2}}}{\Delta B_{i+\frac{1}{2}} + \Delta B_{i-\frac{1}{2}}} & \text{if } \Delta B_{i+\frac{1}{2}} \Delta B_{i-\frac{1}{2}} > 0 \\ 0 & \text{otherwise} \end{cases}, \quad (13)$$

where $\Delta B_{i \pm \frac{1}{2}} = \pm(B_{i \pm 1}^n - B_i^n)$.

The diffusion term ($R_i$) at the $i$th grid point for the $n$th time step is constructed as shown below

$$R_i = \frac{\eta(B_{i+1}^n - B_i^n) - \eta(B_i^n - B_{i-1}^n)}{\Delta \lambda^2} - \frac{B_i^n}{\tau} + S_i, \quad (14)$$

where $\eta$ and $\tau$ represent the diffusion coefficient and decay rate as described in Section 1.1. To calculate the evolution of B due to the diffusion term ($R_i$) in the 1D SFT equation, we have used the tri-diagonal solver. The tri-diagonal matrix inversions are performed through the Thomas algorithm (W. H. Press et al. 1992). This method is second-order accurate in space throughout the computational domain.

We have employed the Neumann boundary condition with a zero value at the latitude boundaries, ensuring unimpeded plasma outflow through these boundaries. The selection of the boundary condition is determined by the physical model, which suggests that the magnetic flux will be propelled toward the poles by meridional flows, eventually accumulating and playing a role in polarity inversion during the subsequent cycle. In the case of the 2D SFT equation (Equation (1)), we implemented periodic boundary conditions along the longitudinal axis.

### 2.2. Scheme 2: Runge–Kutta Implicit–Explicit

RK-IMEX is a second-order technique (in space and time) that evolves the PDE's advection term using an upwind Godunov-type explicit approach and solves the diffusion part implicitly (L. Pareschi & G. Russo 2005; S. Kundu et al. 2021). The advection term $A_i$ is calculated using the advection flux as described in the Explicit scheme (see Equations (9)–(13)). The diffusion term $R_i$ is calculated as described in Equation (14). Similar to the Explicit case, $\Delta t$ is selected using the CFL condition. To evolve $B_i^n$ by time step $\Delta t$ to get $B_i^{n+1}$, we use the following algorithm,

$$B_i^{(1)} = B_i^n + \Delta t \gamma R_i^{(1)}, \quad (15)$$

$$B_i^{(2)} = B_i^n + \Delta t(-A_i^{(1)} + (1 - 2\gamma) R_i^{(1)} + \gamma R_i^{(2)}), \quad (16)$$

$$B_i^{(n+1)} = B_i^n + \frac{\Delta t}{2}(-A_i^{(1)} - A_i^{(2)} + R_i^{(1)} + R_i^{(2)}), \quad (17)$$

where

$$\gamma = 1 - \frac{1}{\sqrt{2}}.$$

The first step is an implicit update of the diffusion term using a tri-diagonal solver. In the second step, the advection flux is explicitly calculated using the previously obtained state $B_i^{(1)}$, while the diffusion term is treated implicitly. In the final step, the state of the next time step is calculated explicitly using the states already obtained $B_i^{(1)}$ and $B_i^{(2)}$. The boundary conditions are similar to that mentioned in the Explicit scheme (Section 2.1).

To advance the evolution of the magnetic field in both $\lambda$ and $\phi$ directions using the 2D SFT equation (Equation (1)), we employ the dimension-splitting technique (E. F. Toro 1999; D. Gąsiorowski 2012) in both schemes (Explicit and RK-IMEX). This allows the use of the 1D algorithms described above to solve the 2D equation where the fundamental equation can be expressed as a linear sum of operators in each direction independently. The 2D SFT equation (Equation (1)) can be written in the form

$$\frac{\partial B_r}{\partial t} + L^\lambda(B_r) + L^\phi(B_r) = 0, \qquad (18)$$

where $L^\lambda$ and $L^\phi$ represent the operations with $\lambda$ and $\phi$, respectively. The numerical update is performed using dimensional splitting and the Strang operator (G. Strang 1968), as shown below:

$$B_r^{n^*} = B_r^n + \frac{\Delta t}{2} \cdot L^\lambda(B_r^n), \qquad (19)$$

$$B_r^{n^{**}} = B_r^{n^*} + \Delta t \cdot L^\phi(B_r^{n^*}), \qquad (20)$$

$$B_r^{n+1} = B_r^{n^{**}} + \frac{\Delta t}{2} \cdot L^\lambda(B_r^{n^{**}}). \qquad (21)$$

In the first step, the magnetic field is updated for $L^\lambda$ over a duration of $\frac{\Delta t}{2}$ to acquire $B_r^{n^*}$. In the subsequent step, $L^\phi$ is updated for a full time step of $\Delta t$, to obtain $B_r^{n^{**}}$. Finally, to complete one full time step update, the magnetic field is once again updated for $L^\lambda$ over a duration of $\frac{\Delta t}{2}$. The Strang method is employed to ensure a second-order accurate temporal evolution even when extending the individual 1D second-order algorithms to 2D calculations.

## 3. Physics-informed Neural Networks for SFT

PINNs are universal function approximators capable of incorporating the governing physical laws, represented by PDEs, into the learning process (M. Raissi et al. 2019). This is achieved by including the PDEs along with the initial and boundary conditions of the system into the loss function, to optimize the underlying neural networks. The loss function is modified to include separate components for the loss from initial conditions ($\xi_{ic}$), losses from boundary conditions ($\xi_{bc}$), and losses from the SFT equation ($\xi_{pde}$). This ensures that the network learns all these physical constraints along with the available data. It is worth noting that the specific problem addressed in this paper does not contain any observed data for training and hence the learning process includes only the physical constraints. The architecture and optimization of hyperparameters are described in the following sections.

### 3.1. Architecture

Figure 1 illustrates the general architecture of PINNs, which includes the input layer (pink color), hidden layers (red), and output layer (cyan) connected through weights (black solid arrows) and biases (M. Raissi et al. 2019). The input layer comprises the latitude ($\lambda$), longitude ($\phi$), and time ($t$), while the output predicts the magnetic field ($B$) at the specified ($\lambda$, $\phi$) coordinates at the given time $t$. Hidden layers consist of fully connected neural networks, forming the overall architecture of the PINN model. Hyperbolic tangent (tanh) is used as the activation function $f(z)$. Here, $z$ represents the weighted sum of values from the preceding neural network layer along with the bias term (I. Goodfellow et al. 2016). The training starts by randomly setting the weights and biases for each connection in the network. For every training point, the network computes the output magnetic field $B$ using PINNs. The loss function is then computed based on satisfying the governing equation (SFT equation), along with initial and boundary conditions. The weights are further updated through backpropagation to minimize the value of the loss function. This iterative process continues until the loss function stabilizes.

In PINNs, the computation of derivatives relies on the automatic differentiation method utilizing the network's structure (weights, biases, and activation function). The outputs are linked to the inputs through activation functions, which have known derivatives. We can determine the derivative of the output ($B$) with respect to the inputs ($t$, $\phi$, and $\lambda$) by applying the chain rule. This makes the PINN method independent of computational grids and allows derivative calculations at any point in the domain. This meshless characteristic of the PINN method is a major advantage, effectively addressing the concerns related to resolution dependence commonly being observed in numerical schemes described in Sections 2.1 and 2.2 above. The absence of any mesh significantly reduces the artificial-intelligence-based algorithm's vulnerability to numerical diffusion (Q. Hou et al. 2023).

The hyperparameter set ($\Theta$) that includes the network depth and width, number of training points, optimizer, learning rates, and iterations is optimized as described in Section 3.2. In this work, we have employed both the ADAM optimizer (D. P. Kingma & J. Ba 2014) and the L-BFGS optimizer (D. C. Liu & J. Nocedal 1989). ADAM is applied first, followed by L-BFGS. This sequential approach takes advantage of each optimizer's strengths: ADAM provides rapid initial convergence, while L-BFGS ensures precise final optimization. For a given set $\Theta$, the loss function used for the learning process is calculated as:

$$\xi(\Theta) = w_{ic}\xi_{ic}(\Theta) + w_{bc}\xi_{bc}(\Theta) + w_{pde}\xi_{pde}(\Theta), \qquad (22)$$

$$\xi_{ic}(\Theta) = \frac{1}{N_{ic}}\sum_{i=1}^{N_{ic}}|B_{pinn}(\lambda^i, \phi^i, t=0; \Theta) - B_{init}(\lambda^i, \phi^i)|^2, \qquad (23)$$

$$\xi_{bc}(\Theta) = \frac{1}{N_{bc}}\sum_{i=1}^{N_{bc}}|B_{pinn}(\lambda_{bc}^i, \phi_{bc}^i, t^i; \Theta) - B_{bc}(\lambda_{bc}^i, \phi_{bc}^i, t^i)|^2, \qquad (24)$$

$$\xi_{pde}(\Theta) = \frac{1}{N_{pde}}\sum_{i=1}^{N_{pde}}|f(\lambda^i, \phi^i, t^i; \Theta)|^2, \qquad (25)$$

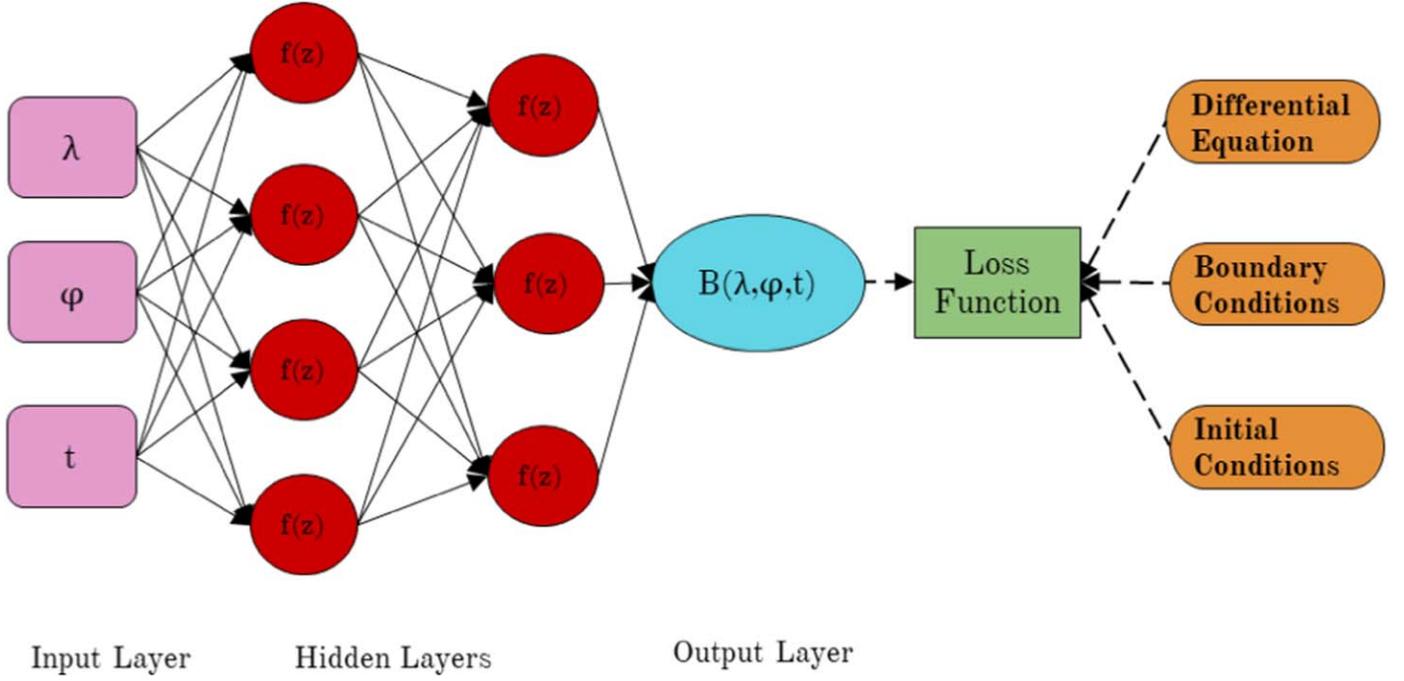

**Figure 1.** Schematic of PINNs to solve SFT equations. The input layer comprises the latitude ($\lambda$), longitude ($\phi$), and time ($t$). The output corresponds to the magnetic field at the specified coordinate and time. The loss function incorporates the physical equation (Equation (1)) and initial and boundary conditions. Note that the hidden layers depicted in the image are for representational purposes only, with actual values specified in Table 1.

where $f(\lambda, \phi, t; \Theta)$ is the residual of the SFT equation given by

$$f(\lambda, \phi, t; \Theta) = \frac{\partial B_{\text{pinn}}}{\partial t} + \frac{1}{R\cos\lambda}\frac{\partial}{\partial\lambda}(B_{\text{pinn}} u(\lambda)\cos(\lambda))$$
$$+ \Omega(\lambda)\frac{\partial B_{\text{pinn}}}{\partial \phi} - \frac{\eta}{R^2 \cos^2\lambda}\frac{\partial^2 B_{\text{pinn}}}{\partial \phi^2}$$
$$- \frac{\eta}{R^2 \cos\lambda}\frac{\partial}{\partial\lambda}\left(\cos\lambda \frac{\partial B_{\text{pinn}}}{\partial \lambda}\right)$$
$$+ \frac{B_{\text{pinn}}}{\tau} - S(\phi, \lambda, t).$$

Here $B_{\text{pinn}}(\lambda, \phi, t; \Theta)$ is the predicted value from the PINNs at the point in the domain $(\lambda, \phi, t)$. $N_{\text{ic}}$, $N_{\text{bc}}$, $N_{\text{pde}}$, respectively, stand for the number of points selected for the training in the initial, boundary condition, and domain. $B_{\text{init}}(\lambda, \phi)$ denotes the initial condition used in the simulation, whereas $B_{\text{bc}}(\lambda^i_{\text{bc}}, \phi^i_{\text{bc}}, t^i)$ represents boundary conditions at boundary points $\lambda_{\text{bc}}$ and $\phi_{\text{bc}}$. Since the problem has two spatial dimensions, the boundary conditions for both $\lambda$ and $\phi$ have been incorporated separately and subsequently added to get the total loss function for the boundary condition. $w_{\text{ic}}$, $w_{\text{bc}}$, $w_{\text{pde}}$, respectively, are the weights given to each term in the loss function during the training process. Each term in the loss can have a different order of magnitude, which results in unequal training of individual terms. The inclusion of the weight in the loss function forces the network to converge to a unique solution following the given constraints. The mean square error (MSE) has been used as the loss function. Since the 1D SFT equations (Equation (4)) do not include terms in $\phi$, the corresponding input and the loss term ($\xi^\phi_{\text{bc}}$) are omitted while simulating the 1D case.

### 3.2. Optimization of the Model Parameters

Hyperparameter optimization plays a crucial role in assessing the SFT equation using PINNs. The overall optimization of the code for both 1D and 2D has been performed using Gaussian process (GP)-based Bayesian optimization. In this method, an initial set of hyperparameters is assumed to begin the optimization process. The network is optimized to calculate the test and train loss using these values. Once the losses are calculated, a GP-based regressor is employed to predict the next set of hyperparameters using the current state. Hence there is a two-level optimization to obtain the best set of hyperparameters (P. Escapil-Inchauspé & G. A. Ruz 2023). A brief description of the optimization technique used is given in the Appendix. We optimized all the hyperparameters including the number of network layers, the number of nodes in a layer, the number of points used for training (initial, domain, boundaries), the iterations required in each optimizer, the learning rate, the activation function, and the weights for the loss function (domain, boundaries, and initial). A hyperparameter list for both 1D and 2D simulations obtained after the optimization is given in Table 1. It is worth noting that in the case of 1D simulation, the SFT parameters described in the introduction part are used along with the source described in the 1D section later. For the 2D case, we take a BMR as described in case 2D. We tested the code for different initial conditions and found that the hyperparameter set is consistent. The same has been verified using multiple simulations with different initial positions, orientations, and polarity reversed.

To comprehend the relationship between the size of the training data set and the model, several training sessions, each with a varying number of training points, have been carried out. The L1 error for both the test and train sets, corresponding to various sizes of the training data set for the 2D case, is

Table 1
Hyperparameters Used in 1D and 2D Simulations

| Parameters | | 1D | | 2D | |
|---|---|---|---|---|---|
| Hidden layers | [nodes] × layers | [41] × 10 | | [32] × 8 | |
| Number of training points | Initial ($N_{ic}$) | 2787 | | 28838 | |
| | Boundary ($N_{bc}$) | 2356 | | 20022 | |
| | Domain/PDE ($N_{pde}$) | 87460 | | 173567 | |
| Optimizer | | ADAM | L-BFGS | ADAM | L-BFGS |
| Weights on the loss | Initial ($w_{ic}$) | 0.60 | 0.32 | 0.17 | 0.25 |
| | Boundary $\lambda$ ($w_{bc}^{\lambda}$) | 0.01 | 0.04 | 0.16 | 0.25 |
| | Boundary $\phi$ ($w_{bc}^{\phi}$) | - | - | 0.45 | 0.25 |
| | Domain/ PDE ($w_{pde}$) | 0.39 | 0.64 | 0.22 | 0.25 |
| Learning rate | | 0.0022 | - | 0.00052 | - |
| Iterations | | 35016 | 14166 | 199495 | 7904 |

**Note.** This table describes the hyperparameter set including the hidden layer configuration, number of training points for different phases and optimizer choices. The values are obtained using the GP-based Bayesian optimization. In both cases, training is conducted in two steps: first with ADAM, and then with L-BFGS. Weights assigned to loss components, learning rates, and iterations for both optimization steps are also mentioned.

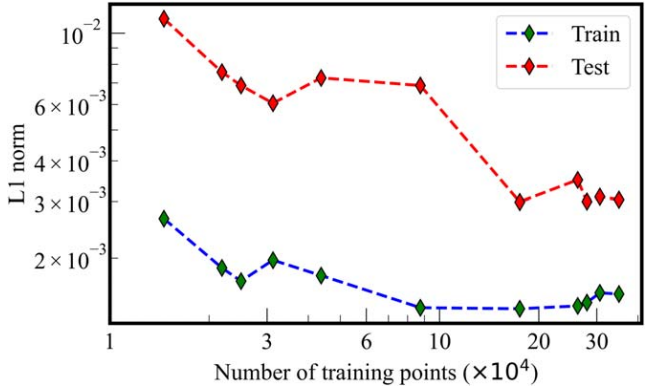

**Figure 2.** Variation in the error (L1 norm) on both the test and training sets on varying the number of training points. The error reduces initially and gets saturated.

illustrated in Figure. 2. Despite initial fluctuations, the errors stabilize after $\sim 2 \times 10^5$ training points for both test and training data sets. This observation serves as a confirmation that the model optimization is successful and that the provided architecture consistently yields minimal errors.

## 4. Results: Validation

Validation of the codes is carried out in two ways. Initially, we verified the numerical (RK-IMEX and Explicit) and PINN codes using the analytical form proposed by C. R. DeVore et al. (1984). Also, the numerical codes were further validated by evolving BMRs and comparing the outcomes with the approach described by A. R. Yeates (2020).

### 4.1. Analytical Validation

To validate the numerical (RK-IMEX and Explicit) and the PINN methods, we use the analytical solution prescribed by C. R. DeVore et al. (1984). The analytical method takes the meridional flow and diffusion terms in the 1D SFT equation into account (first and second terms on the right-hand side of Equation (4)). The relative effect of meridional flow and diffusion is represented by ($\beta = \frac{u_0 R_\odot}{\eta}$). We use $\beta = 10$ for our validation following the range of values adopted by Leighton (R. B. Leighton 1964; C. R. DeVore et al. 1984).

The model assumes an initial state $B(\mu, t=0)$ as shown below

$$B(\mu, t=0) = B_0 \, \beta \exp(-\beta) \sinh(\beta\mu), \quad (26)$$

where $\mu = \cos\theta$ and $\theta = 90° - \lambda$. A special advection profile $u(\theta)$ is considered for this specific case given by

$$u(\theta) = -u_0 \sin\theta \tanh\left(\frac{\beta}{2}\cos\theta\right). \quad (27)$$

Here $u_0 = 12.5 \text{ m s}^{-1}$ denotes the amplitude of the velocity profile. The diffusion coefficient $\eta = 500 \text{ km}^2 \text{ s}^{-1}$ is used as described in Section 1.1. This gives an exponentially decaying analytical solution as given below

$$B(\mu, t) = B(\mu)\exp\left(-\frac{2\beta^2 e^{-\beta} t}{\tau_d}\right) \quad (28)$$

with decay time constant $\tau_d = \frac{R_\odot^2}{\eta}$.

In Figure 3, we compare the results obtained in the PINNs and RK-IMEX with the analytical solution. The absolute error between the PINNs and analytical solution is in the range (0.2%–1.3%). As expected for any numerical scheme, the RK-IMEX method shows higher accuracy in the analytical solution on increasing the number of grid points (Ngrid; see Figure 3(b)). This is further supported by the error plot, where the L1 norm demonstrates a decreasing trend while increasing Ngrid from 256 to 2048 bins. PINNs exhibit the lowest error even compared to the highest Ngrid simulation using the RK-IMEX scheme (Figure 3(a)). Because of its mesh independence, the PINN solution exhibits fewer fluctuations in the error.

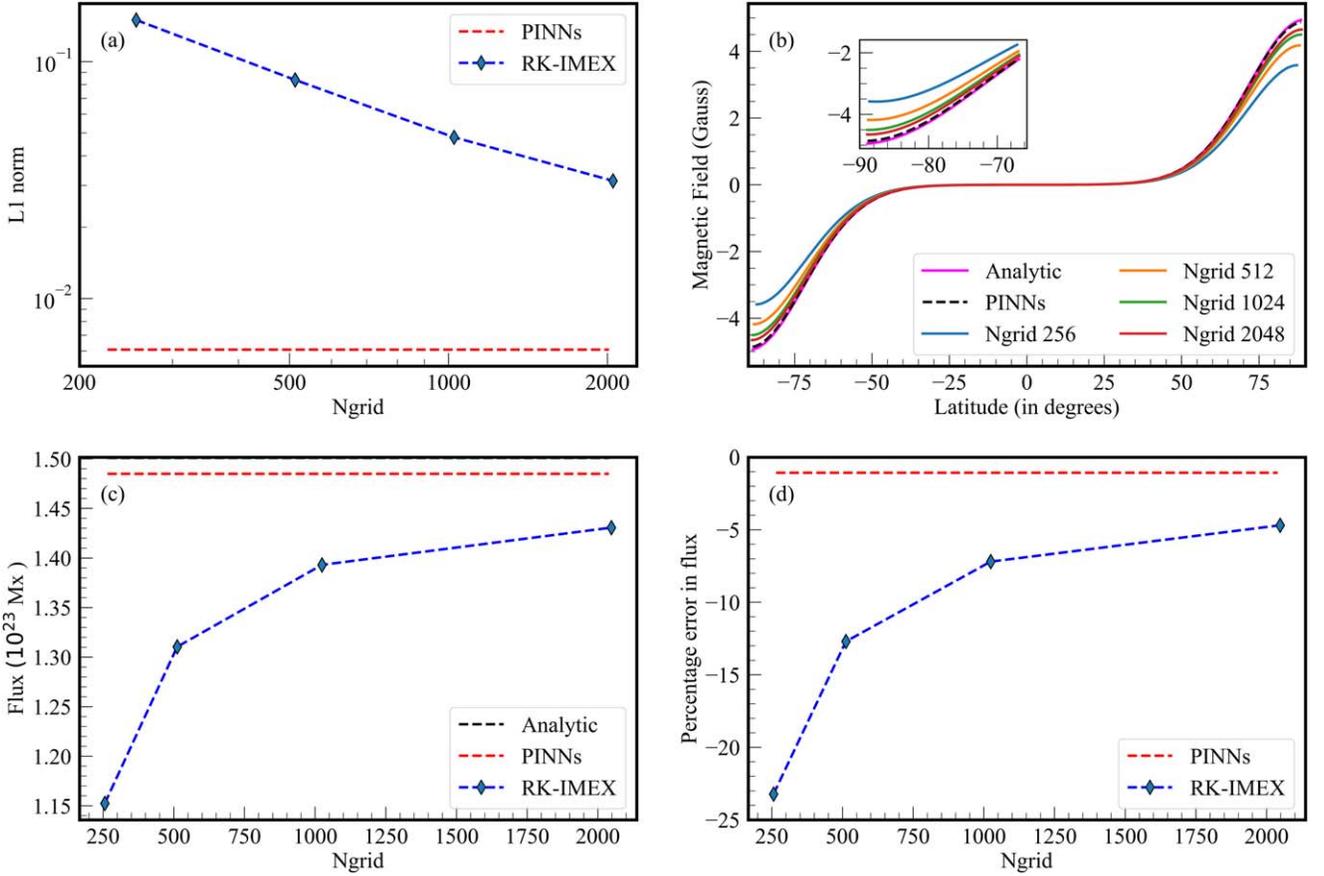

**Figure 3.** Evolved magnetic field for $\beta = 10$. (a) L1 norm calculated for the PINNs and RK-IMEX scheme compared to the analytic solution. (b) Solution for the time stamp $t = \tau_d$ for different numbers of grid points. The PINN solution (blue) and analytical solution (black dashes) are also plotted. (c) Total flux in the southern hemisphere. (d) Percentage error of the flux with respect to the analytical flux.

The total flux is calculated as

$$\Phi(t) = 2\pi R^2 \int_{-1}^{1} B(\mu, t) d\mu. \quad (29)$$

The comparison of flux calculated for the southern hemisphere is plotted in Figure 3(c). It is observed that the flux loss decreases on increasing Ngrid. The error in flux from the PINN solution is ~1.07%, which is lower than that of the RK-IMEX method with the highest Ngrid (see Figure 3(d)). The PINN solution produces the best match for the flux compared to the numerical method. This shows that the flux loss in the numerical method is higher than in the PINN method. The flux loss could critically affect the solutions produced using SFT simulations, especially when the aim is to calculate the polar flux at the end of a solar cycle.

### 4.2. Sensitivity Test

To verify the stability of the chosen hyperparameters, we performed a sensitivity test (ST). This involved randomly altering some of the hyperparameters and retraining the network. Table 2 illustrates the perturbed hyperparameters and their resulting magnetic and flux errors for various sets. The change in hyperparameters is expressed in percentages, where an increase (decrease) is indicated by ↑ (↓). The magnetic field error was found to be high in all the sets in comparison to that of the optimized hyperparameter set (error = 0.61%) (Figure 4(a)). We found that the flux errors range from 1% to 14% as depicted in Figure 4(b). The flux errors are lower in some of the sets (ST1,4,6,8 and 10), although corresponding errors in the magnetic field are higher (as compared to the error for the optimized set of hyperparameters). This exercise indicates the consistency and stability of the selected hyperparameter set.

### 4.3. Evolution of BMRs

To validate the numerical (RK-IMEX and Explicit) and PINN methods, we evolve the BMRs after initializing at various latitudes (0° and 12°) as shown in Figures 5(a) and (d). We utilized a resolution of $\Delta\lambda = \Delta\phi = 1°$ for our simulations. Once the BMR is initialized, the average magnetic field is calculated along the longitude to make the 1D magnetic profile (Figures 5 (b) and (e)). The initial magnetic field distribution in latitude ($\lambda$) is fitted using the function given below

$$B(\lambda) = \left[ e^{-0.5\left(\frac{\lambda - \lambda_c}{\sigma_c}\right)^2} + e^{-0.5\left(\frac{\lambda + \lambda_c}{\sigma_c}\right)^2} \right] [A \sin(B\lambda + C) + D]. \quad (30)$$

The fitted parameters are $A = 14.88$, $B = 0.14$, $C = -3.65$, $D = -7.07$, $\lambda_c = -2.33$, and $\sigma_c = 3.51$ for the first case (0°), and $A = -14.88$, $B = 0.14$, $C = -2.24$, $D = -7.07$, $\lambda_c = -0.67$, and $\sigma_c = 3.51$ for the second case (12°). This distribution is then evolved using the 1D SFT equation without the source term. The BMR and parameters are chosen from A. R. Yeates (2020). The evolved magnetic field, along

Table 2
Different Sets of Perturbed Hyperparameters, with the Percentage Change in Hyperparameters (↑ : Increase and ↓ : Decrease) and the Corresponding Percentage Errors in the Magnetic Field and Total Flux in a Hemisphere

| ST no. | Hyperparameter | % change in hyperparameter | % error in magnetic field | % error in total flux |
|---|---|---|---|---|
| 1 | $N_{ic}$ | 20 ↓ | 1.05 | 1.25 |
| 2 | $N_{ic}$ | 20 ↑ | 3.52 | 5.15 |
| 3 | nodes | 25 ↓ | 2.79 | 2.10 |
| 4 | layers | 25 ↓ | 0.66 | 1.23 |
| 5 | layers | 25 ↑ | 3.27 | 2.19 |
| 6 | $N_{bc}$ | 50 ↓ | 1.15 | 0.97 |
| 7 | $w_{ic}$ | 18 ↓ | 0.96 | 2.05 |
| 8 | layers<br>Learning Rate | 50 ↓<br>100 ↑ | 1.55 | 0.91 |
| 9 | nodes<br>$w_{ic}$ | 25 ↓<br>18 ↓ | 6.03 | 14.60 |
| 10 | nodes<br>Learning Rate | 25 ↓<br>100 ↑ | 3.65 | 0.58 |
| 11 | nodes<br>layers | 25 ↓<br>25 ↓ | 1.30 | 2.65 |

**Note.** The errors are calculated with respect to the analytic solution.

with the respective initial states and BMRs are plotted in Figures 5(c) and (f).

When the BMR is located at the equator (0°), each magnetic polarity is carried away in opposite directions, resulting in distinct polarities at each pole. The flux cancellation is very low and the flux loss is mainly due to the diffusive and decay losses. When the BMR is initiated in the northern hemisphere (12°), the magnetic field mostly gets pushed toward the corresponding pole. The magnetic field intensity gets reduced due to the flux cancellations between different polarities advecting together. There is a very small amount of field that has crossed the equatorial region because of the initial diffusion. The solutions from the RK-IMEX and PINN schemes obtained in the present study are compared with the standard numerical scheme from A. R. Yeates et al. (2023); hereafter referred to as Yeates scheme (see Figure 6 in their paper). We observe that the features obtained in the evolution of BMR $<B_r(t)>$ are similar to columns 3 and 4 of Figure 5 of this work. We have further estimated the L1 norm obtained between the solution from both the schemes (RK-IMEX and PINN) and that of the Yeates scheme. For cases 1 and 2, the difference is 9% and 14%, respectively, between the RK-IMEX and Yeates schemes. Such a deviation is expected as the Yeates scheme is first-order accurate, whereas the RK-IMEX scheme has second-order accuracy. The difference between the PINN method and the Yeates scheme is 11% and 15% for cases 1 and 2, respectively. Additionally, the difference between the RK-IMEX and PINNs method is found to be ∼6% in both cases 1 and 2. This is expected because of the lesser numerical diffusion in the PINN method compared to the other two numerical methods.

## 5. Results: Case Studies

After the validation process, the codes were employed to examine various scenarios as follows:

1. A 1D SFT simulation that integrates a source term exploring the evolution of magnetic field with respect to latitude (Section 5.1).
2. A 2D SFT simulation that introduces a BMR at a specific coordinate in the $\theta$–$\phi$ plane. This aims to understand the spatial distribution and evolution of magnetic fields in two dimensions (Section 5.2).
3. An analysis based on a BMR generated from magnetogram data. This case study uses real observational data from the Michelson Doppler Imager (MDI) aboard the Solar and Heliospheric Observatory (SOHO) to validate the model's accuracy in reproducing the magnetic field dynamics observed on the solar surface (Section 5.3).

### 5.1. Case Study: 1D

We evolved the magnetic field using the 1D SFT equation (Equation (4)) by including the meridional flow, diffusion, decay, and source terms using both numerical schemes (Explicit and RK-IMEX) and PINNs. The parameters used for advection, diffusion, and decay are described in Section 1.1. The source term $S(\lambda, t)$ represents a smooth distribution that characterizes the probability of preceding and following polarities emerging on the solar surface. This representation is akin to the approach employed by M. Dikpati et al. (2006) and consists of a pair of rings featuring opposite magnetic polarities. The path of the new ARs followed by a complete cycle ($\Lambda_0$) is determined by the quadratic approximation established based on extensive observations of solar cycles (J. Jiang et al. 2011).

$$\Lambda_0 = 26.4 - 34.2\left(\frac{t}{P}\right) + 16.1\left(\frac{t}{P}\right)^2 \quad \text{(in degrees)}, \quad (31)$$

where $P = 11$ yr is the time period of a typical solar cycle. The source term is given by the equation

$$\begin{aligned}
S(\lambda, t) = &\ kA_m S(t) S_0[\lambda; \Lambda_0(t) - \Delta\lambda(t), \delta\lambda] \\
&- kA_m S(t) S_0[\lambda; \Lambda_0(t) + \Delta\lambda(t), \delta\lambda] \\
&+ kA_m S(t) S_0[\lambda; -\Lambda_0(t) - \Delta\lambda(t), \delta\lambda] \\
&- kA_m S(t) S_0[\lambda; -\Lambda_0(t) + \Delta\lambda(t), \delta\lambda],
\end{aligned} \quad (32)$$

where $k = \pm 1$ is used alternating between even and odd cycles and $A_m = 0.003$ is the amplitude of the source term (K. Petrovay & M. Talafha 2019). The time profile $S(t)$ of solar activity in a typical cycle was determined from the average of many cycles as (D. H. Hathaway et al. 1994)

$$S(t) = \frac{at_c^3}{\exp\left(\frac{t_c^2}{b^2}\right) - c} \quad (33)$$

with $a = 0.00185$, $b = 48.7$, $c = 0.71$, and $t_c$ representing the time since the last cycle minima. The latitudinal profile

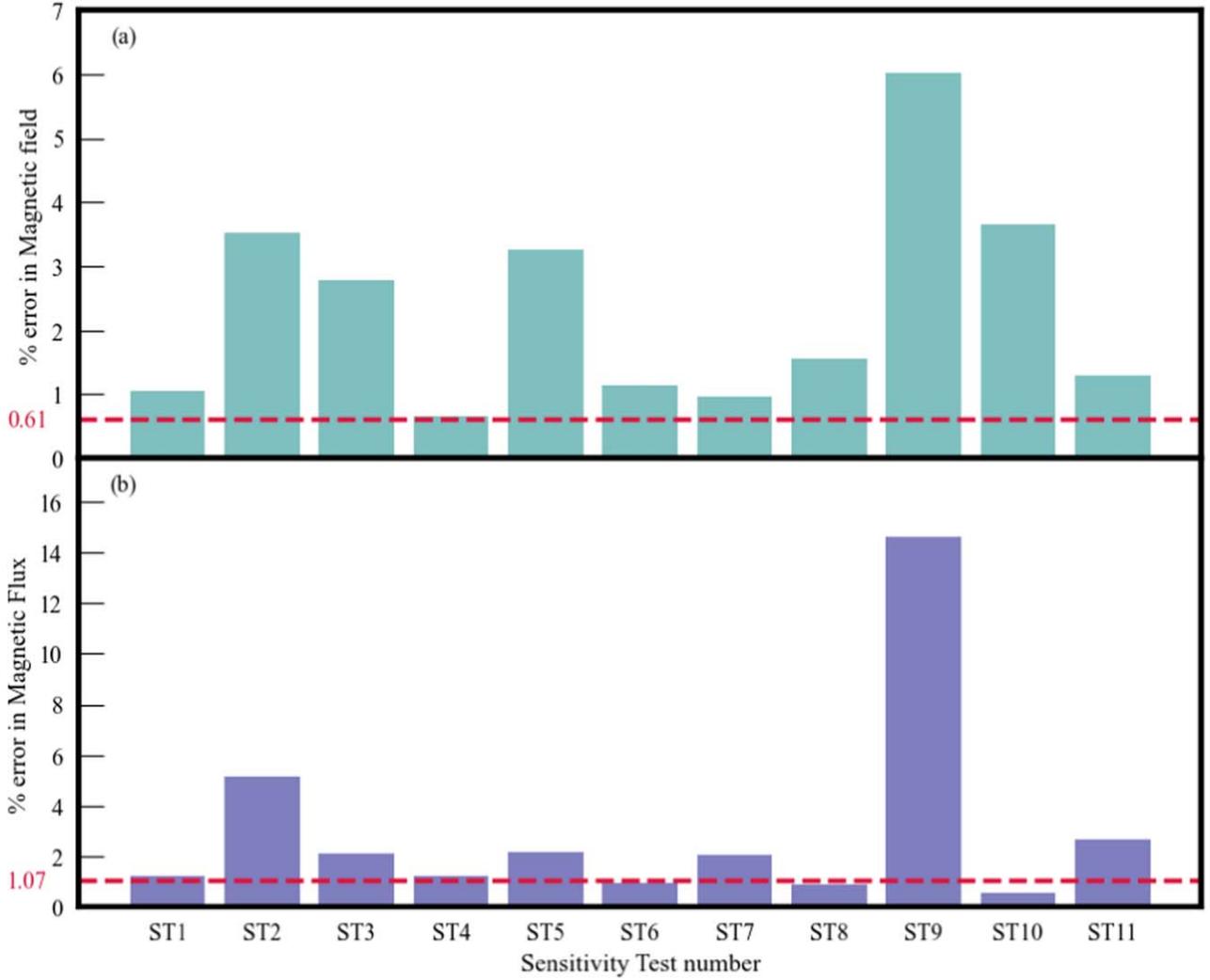

**Figure 4.** Percentage error in (a) magnetic field and (b) magnetic flux due to different sets of hyperparameter perturbation for the analytic validation as given in Table 2. The horizontal dashed lines correspond to the error (magnetic field and flux) for the optimum hyperparameters used as given in Table 1.

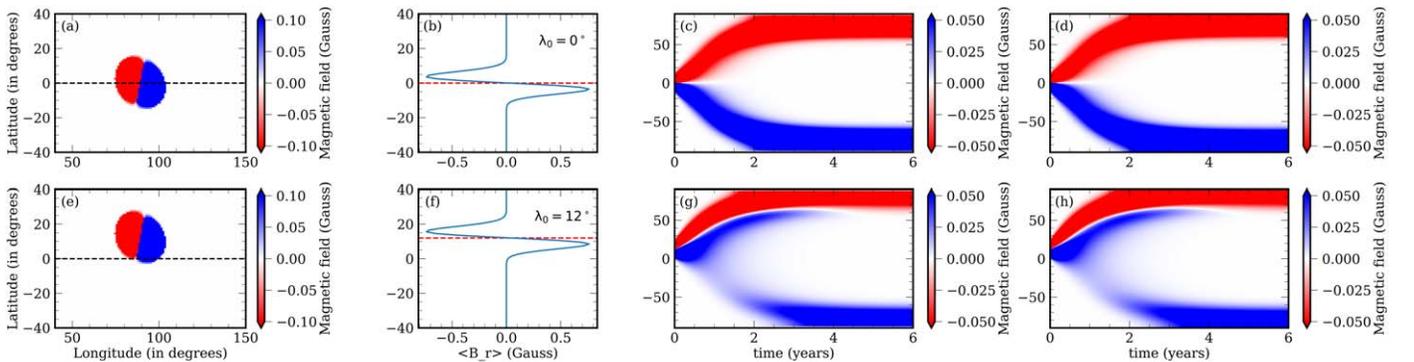

**Figure 5.** Evolution of BMRs using the RK-IMEX at different initial latitudes (row 1: 0° and row 2: 12°). Column 1: initial BMR planted in the $\theta$–$\phi$ plane $B_r(t=0)$. Column 2: magnetic field averaged in longitude $<B_r(t=0)>$. Columns 3 and 4: average magnetic field obtained through the evolution of BMR using 1D SFT equation $<B_r(t)>$ with RK-IMEX (column 3) and PINN (column 4) methods.

$S_0(\lambda; \Lambda_0, \delta\lambda)$ is a Gaussian given by

$$S_0(\lambda; \Lambda_0, \delta\lambda) = \exp\left(\frac{-(\lambda - \Lambda_0)^2}{2(\delta\lambda)^2}\right), \quad (34)$$

where the full width at half maximum $\delta\lambda = 6°$. As the solar cycle progresses, sunspots tend to appear at higher latitudes during the cycle's peak and migrate toward lower latitudes as the cycle declines. This phenomenon contributes to the latitudinal separation ($\Delta\lambda$) observed in the rings and is a result

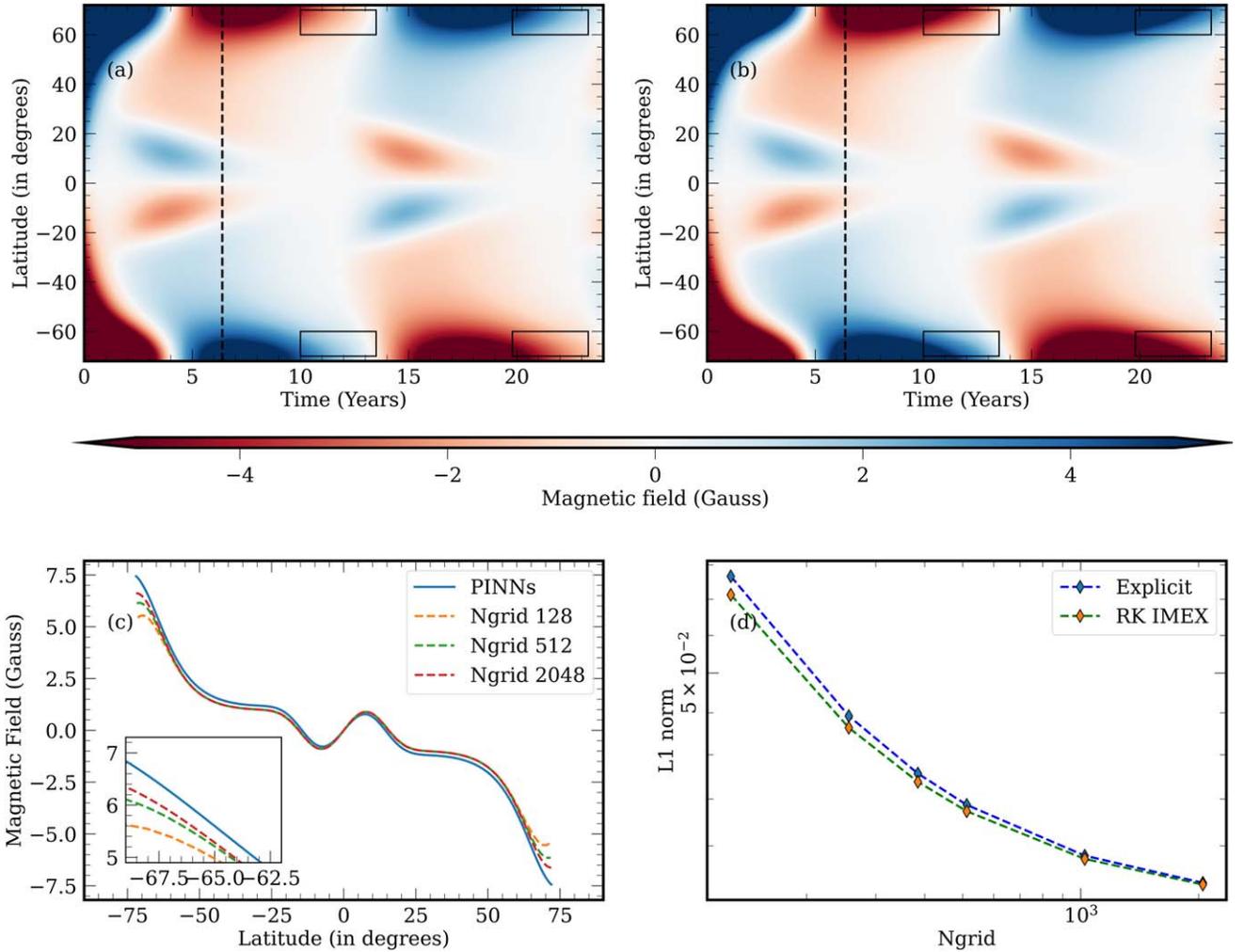

**Figure 6.** Row 1: butterfly diagram obtained by solving the 1D SFT equation including the source term. (a) numerical solution obtained with the RK-IMEX scheme. The solution for the Explicit method is not plotted since it is similar to the RK-IMEX scheme. (b) solution attained through the PINN network. The rectangular box marks the differences near the boundary between the two methods. Row 2: different resolutions of numerical solutions compared with the PINN solution. We can see a decrease in error with increasing resolution. (c) solution for $t = 6.384$ yr (corresponding to the vertical lines in (a) and (b)). The variation in the boundary can be seen in the inner plot. (d) variation in the loss with an increase in the number of grid points (Ngrid) for both numerical schemes (RK-IMEX and Explicit). The L1 norm is calculated by comparing the numerical solutions with the PINN solution.

of Joy's law as shown below.

$$\Delta\lambda = 0.25 \frac{\sin\lambda}{\sin 20°}. \quad (35)$$

We defined our simulation box to span from $-70°$ to $+70°$, excluding the polar regions, to avoid the singularity associated with magnetic flux and $\cos(\lambda)$ term in the SFT equation. The simulation was carried out for 24 yr, with the initial condition set as $B = B_0 \sin(\lambda)$. We have employed boundary conditions as mentioned in the Explicit scheme (Section 2.1).

The comparison of the solution of the PINNs and both numerical methods (Explicit and RK-IMEX) with the same initial condition and empirical equations are given in Figure 6. The solution obtained using the RK-IMEX (Ngrid=2048) and PINN schemes are plotted in Figures 6(a) and (b), respectively. We have also obtained solutions for Ngrid values 128 and 512 using RK-IMEX. Figure 6(c) illustrates the outcomes achieved by PINNs and RK-IMEX for different values of Ngrid at a particular time instance ($t = 6.384$ yr). This temporal point is indicated by black dashed lines in Figures 6(a) and (b). For the sake of brevity, we have not plotted the solution obtained from the Explicit scheme as it is similar to the RK-IMEX solution. The results demonstrate a noticeable alignment between the outcomes obtained through the RK-IMEX and PINN method. However, it is important to recognize that differences exist in the solutions, particularly near the boundary as evident from the inset plot in Figure 6(c). There is a distinct shift in the time of polarity inversion near the poles (see black rectangular boxes in Figures 6 (a) and (b)). The L1 norm of both the numerical schemes (RK-IMEX and Explicit) with respect to the PINN solution is shown in Figure 6(d). By increasing the number of grid points, the numerical solutions for both the RK-IMEX and Explicit schemes gradually converge toward the PINN solution. Both the RK-IMEX and Explicit schemes exhibit a similar order of accuracy, whereas RK-IMEX has a lower error value.

The loss function for training the PINNs for 1D is shown in Figure 7. The weighted loss plots the loss value for which the optimization was performed that includes the weights in each term (see Equation (22) and Table 1). The unweighted loss is calculated after dividing out the weights from the weighted loss. The ADAM optimizer effectively reduces the loss initially by considering the initial magnetic field and the source terms that are given higher weights. The L-BFGS optimizer further minimized the loss of each term to achieve improved accuracy. The loss function tends to stabilize and saturate as the training progresses toward the later iterations, indicating that the optimization process has converged to a satisfactory solution. This convergence assures that the results obtained are reliable and reflect a good level of accuracy in our model.

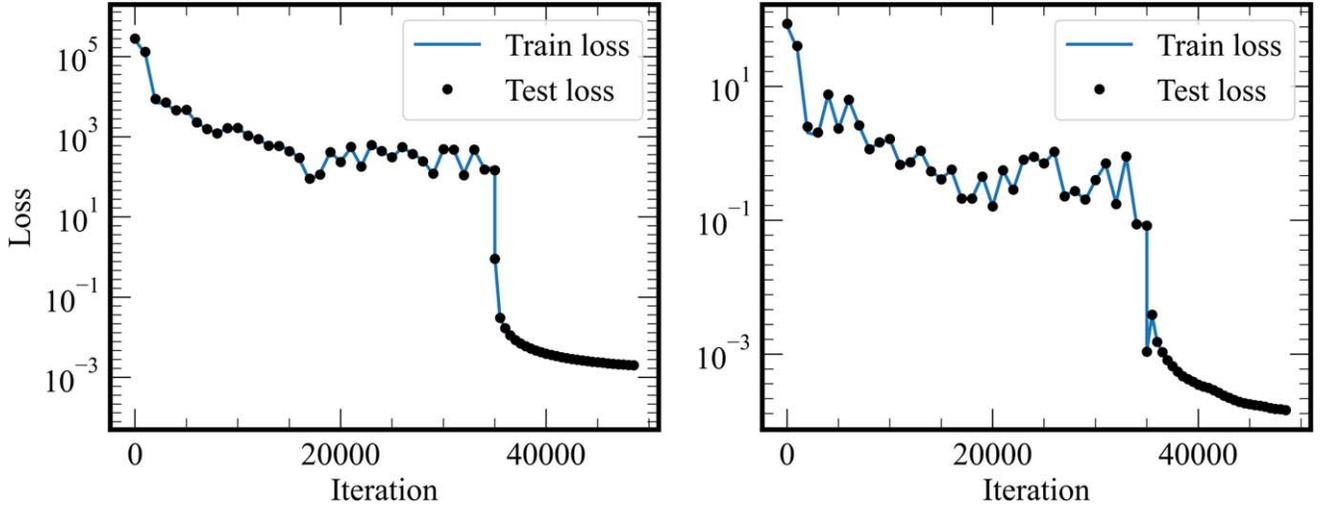

**Figure 7.** Training and test losses for the PINNs for 1D SFT equation. Left: The weighted loss considers the weights that have been applied to each term during training. Right: The unweighted loss is calculated by dividing out the weight giving the actual weight of the training.

### 5.2. Case Study: 2D

We use the numerical (Explicit and RK-IMEX) and PINN methods to evolve a BMR planted at a point on the $\theta$–$\phi$ plane using the 2D SFT equation (Equation (1)). The model parameters including the advection profile (u), diffusion coefficient ($\eta$), and decay constant ($\tau$) are chosen as described in Section 1.1. The initial condition is taken as a BMR with positive and negative circular regions for the respective polarity AR emergence. We follow the prescription by A. A. van Ballegooijen et al. (1998) to artificially introduce this BMR as described in Equations (36)–(39).

$$B_r^\pm = B_0 \exp\left(-2\left(\frac{1 - \cos(\beta_\pm)}{\sigma^2}\right)\right), \quad (36)$$

$$B_r = B_r^+ - B_r^-. \quad (37)$$

Here, $\beta_\pm$ are the heliocentric angles between each point ($\theta = 90° - \lambda$, $\phi$) and the center of each polarity region. $\rho_0$ represents the angular separation between the center of the positive and negative polarities ($\theta_\pm$, $\phi_\pm$).

$$\cos(\beta_\pm) = \cos(\theta_\pm)\cos(\theta) + \sin(\theta_\pm)\sin(\theta)\cos(\phi - \phi_\pm), \quad (38)$$

$$\cos(\rho_0) = \cos(\theta_+)\cos(\theta_-) + \sin(\theta_+)\sin(\theta_-)\cos(\phi_+ - \phi_-). \quad (39)$$

An angular width of $\sigma = m\rho_0$ is assumed around the center of each polarity. For small $\beta_\pm$, the BMR is similar to a Gaussian distribution with this angular width to account for the contribution of the respective polarities. A linear sum of the contribution from each polarity is used as the initial magnetic field distribution. To illustrate, we present results using $m = 0.4$ and ($\theta_0$, $\phi_0$) = (75°, 270°) as parameters for generating the artificial AR. The evolution of this initial state is performed using Equation (1) without including the source term for 110 days. We implemented periodic boundary conditions along the longitudinal axis and outflow conditions along the latitudinal boundaries as described in Section 2.1.

The comparison between the solution obtained from PINNs and the RK-IMEX ($\Delta\lambda = \Delta\phi = 0°.5$) scheme is depicted in Figure 8. The BMR generated using the above parameters is plotted in Figure 8(a). The evolved states for the numerical and PINN methods are shown in Figures 8(b) and (d), respectively. The PINN result exhibits a high degree of similarity to the RK-IMEX solution, indicating strong agreement between the two (Figure 8(c)). The mean relative error between the solutions is $\sim$1% (with mean absolute error < 0.0012 G) for the resolution $\Delta\lambda = \Delta\phi = 0°.5$. The error calculated with respect to the PINN solution is observed to fall on increasing the resolution, similar to the 1D case. Hence, it can be inferred that the numerical solution on increasing the number of grid points approaches the PINN solution.

To assess the additivity of the magnetic field with respect to the SFT equation, we introduced two BMRs at different locations on the $\theta$–$\phi$ plane. This approach is possible because the SFT equation is a linear differential equation (see Section 1.1). The initial seed BMRs (BMR1 and BMR2) are centered at ($\theta_1$, $\phi_1$) = (75°, 270°) and ($\theta_2$, $\phi_2$) = (90°, 234°), respectively (Figure 9(a)). Initially, these BMRs are evolved independently using the PINN method (Figures 9(c) and (e)). Figure 9(d) demonstrates the sum of the individually evolved states. The PINN method is also used to solve the SFT equation for both the BMRs initiated together. The solution obtained by evolving the magnetic field simultaneously is shown in Figure 9(b). These two solutions are then compared to understand the additivity of magnetic field for multiple BMRs. The difference between the solutions is $\sim$1% (Figure 9(f)). This demonstrates that the evolution of multiple BMRs can be effectively accomplished using the PINNs approach by individually evolving each BMR and then summing their contributions. The PINN method is also tested for initial conditions with reversed polarity and has been demonstrated to work effectively.

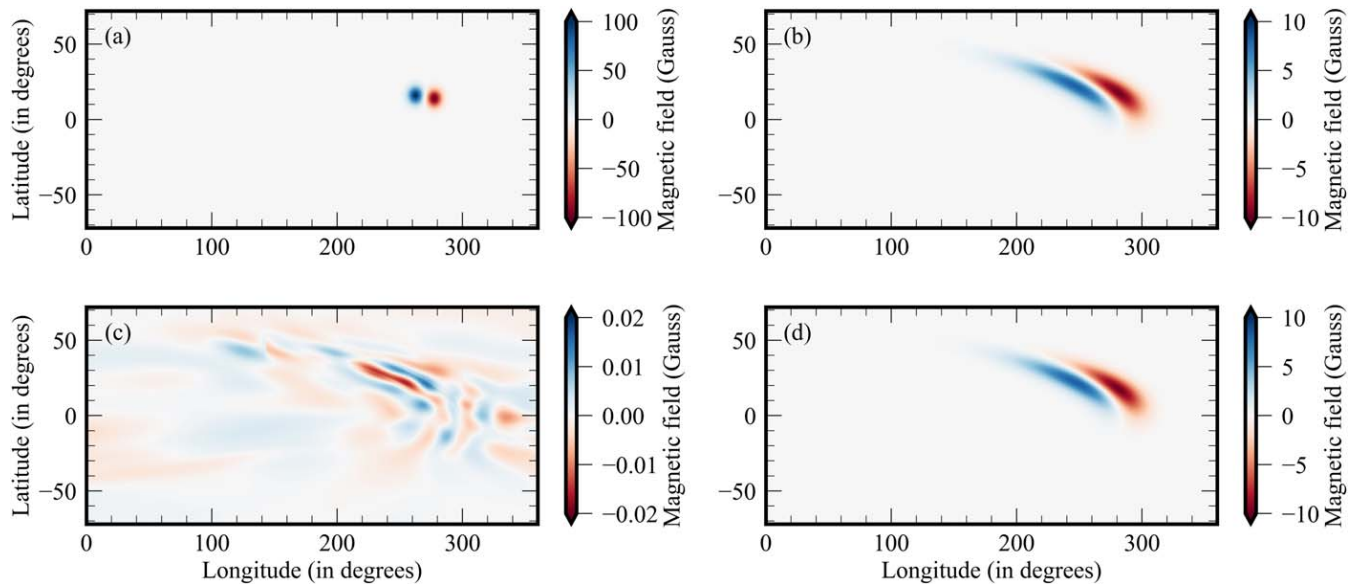

**Figure 8.** Solution for the SFT 2D equation. (a) Initial AR at (75°, 270°), (b) solution using the RK-IMEX scheme, (c) difference in the solutions from the PINN and RK-IMEX schemes, and (d) solution using the PINN method.

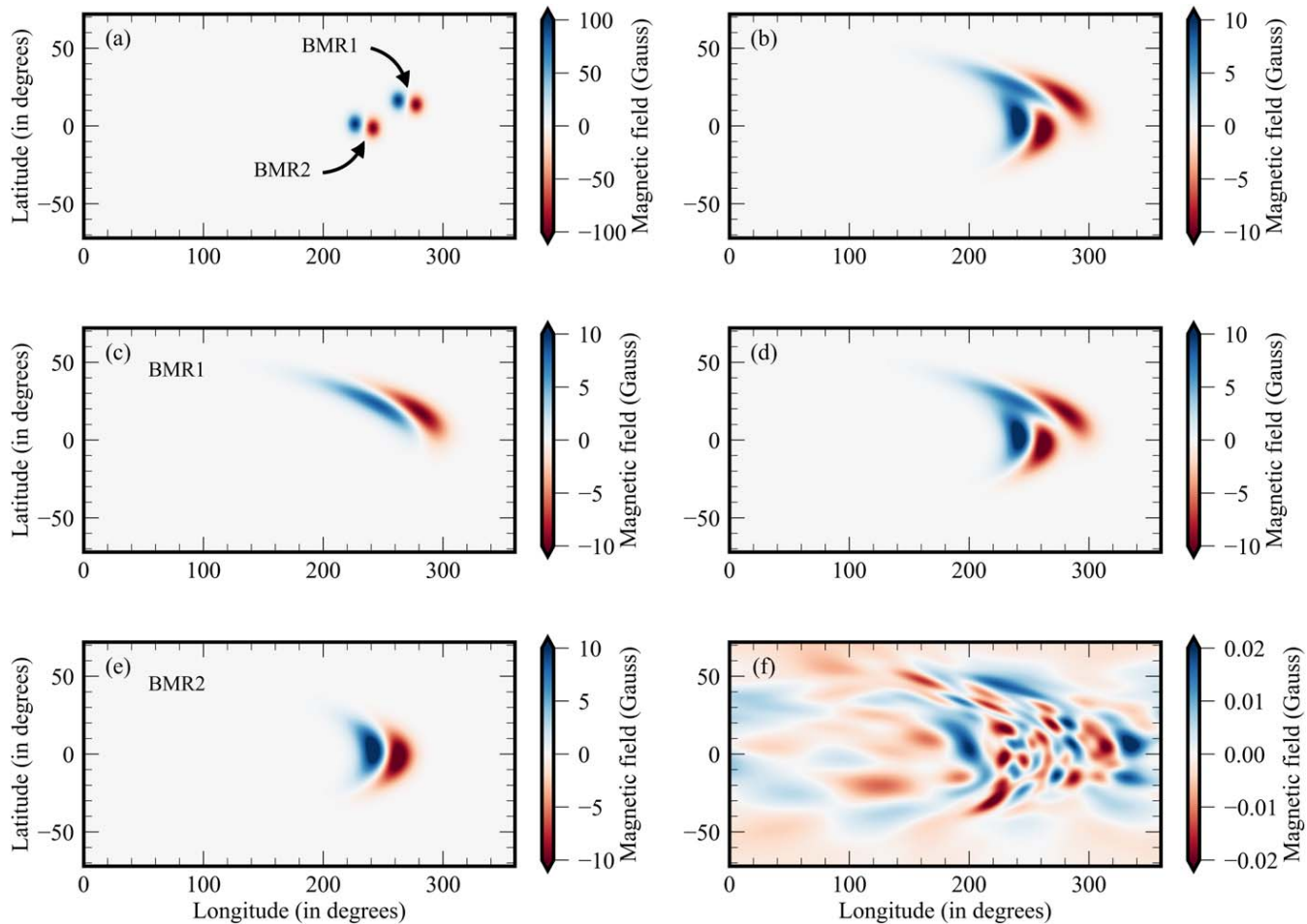

**Figure 9.** Evolution of multiple BMRs using the PINN method. (a) Initial state from which the multiple BMRs (BMR1 and BMR2) evolve. (b) Magnetic field when the BMRs evolved together. (c) Evolved magnetic field of BMR1. (e) Evolution of BMR2. (d) Sum of the individually evolved BMRs. (f) Difference between the sum of BMRs that evolved individually and the BMRs that evolved together.

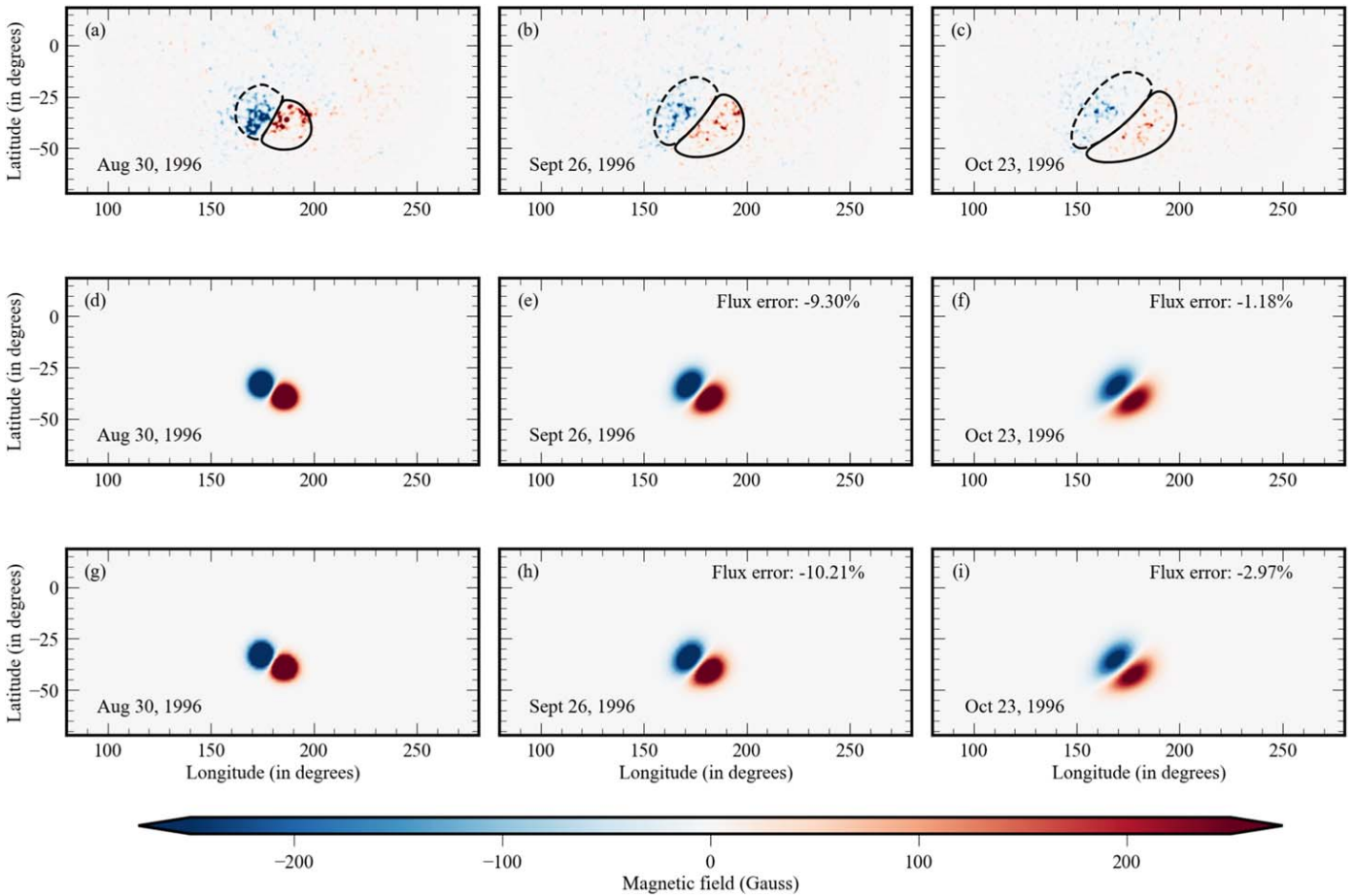

**Figure 10.** Evolution of AR 7978: The top panels ((a), (b), (c)) display data from SOHO/MDI, the middle panels ((d), (e), (f)) demonstrate the evolution of the BMR using the PINN method, and the bottom panels ((g), (h), (i)) showcase solutions obtained through the RK-IMEX method. The columns (left to right) illustrate the magnetic field for three different time stamps: 1996 August 30, September 26, and October 23, respectively. Panels (d) and (g) show the BMR used as the initial condition for both (PINNs and RK-IMEX) methods. To visually assess the alignment between PINNs and observed BMR regions, the BMR region from PINNs is outlined as a contour over the data in the first row. The percentage error in the total unsigned flux compared to the data is mentioned in the top right of the respective panels.

### 5.3. Case Study: Magnetogram Data of AR 7978

The applicability of the models to observed magnetogram data is assessed by evaluating them using data from SOHO-MDI. SOHO is a joint project between the European Space Agency (ESA) and NASA. We have used the "fd_M_96m_01d" data series that consists of the full disk magnetogram of $(1024 \times 1024)$ grids with $(4'')$ resolution in a spectral range of $6767.8 \text{ Å} \pm 190 \text{ mÅ}$ and a cadence of 96 minutes (P. H. Scherrer et al. 1995). AR 7978 is used as an illustration of the application and can be replicated in any other AR. This AR was observed in the southern hemisphere during the solar minima between cycles 22 and 23 (1996). It persisted for multiple CRs (1911–1916), allowing for the examination of its development for at least two rotations after the peak of flux emergence (A. Ortiz et al. 2004; L. van Driel-Gesztelyi & L. M. Green 2015). Moreover, it has distinct north and south polarities, making it an ideal candidate for studying the applicability of the PINN model. To generate the magnetic field of the initial condition, we applied the BMR prescription as described in Equations (36)–(39). The PINN and RK-IMEX methods are used to evolve the BMR over time (see Figure 10). To visually analyze the alignment of PINN's BMR solution and the actual data, the BMR region is marked as contours in the data. The contour traces the magnetic field flow over the data, verifying the evolution. The total unsigned flux is calculated for both the PINN and RK-IMEX solutions with respect to the data. The flux error for PINNs ranges from $\sim 1\%$ to 10%, whereas it is $\sim 3\%$–11% for the RK-IMEX method. The flux error is lower for the PINN method as we obtained in the 2D case. The exact values of the flux error may vary based on the parameters of the SFT modeling and the algorithm used for generating the initial BMR.

### 6. Summary

We demonstrate the applicability of PINNs in solving advection–diffusion equations by simulating SFT equations on the solar surface. The performance of PINNs has been compared with different numerical schemes (Explicit and RK-IMEX). Being scale-independent, the PINs method enhances computational accuracy in simulating solar surface flows by overcoming the challenges posed by conventional grid-based methods. We employed GP-based Bayesian optimization to tune the hyperparameters for the PINN model. The PINN and numerical codes have been validated by comparing their outputs against analytical solutions.

The 1D SFT equation (Equation (4)) with various flow parameters and source terms are solved using the PINN method and subsequently compared to the RK-IMEX solution. We found that the RK-IMEX solution converges to the PINN solution as the number of grids (Ngrids) increases, as depicted by a decrease in the mean absolute error (Figure 6(c)). The accuracy offered by PINNs proves valuable, particularly in achieving precise flux calculations. Its ability to minimize flux loss during advective transport ensures better flux conservation. Estimating the polar magnetic field strength toward the end of the solar cycle demands precise flux values, which are crucial in modeling the strength of the upcoming solar cycles. Improved flux calculations using PINNs can significantly enhance our understanding of the upcoming solar cycle's strength.

In addition to the 1D SFT equation, we explored the applicability of PINNs in understanding the evolution of solar ARs. This is achieved by solving the 2D SFT equation (Equation (1)) to evolve artificially planted BMRs on the $\theta$–$\phi$ plane using both methods (PINNs and RK-IMEX). We found that the mean absolute error between the methods is $\approx 1\%$, highlighting the effectiveness of the PINN method. Once the hyperparameters are tuned, PINNs consistently perform well regardless of the initial position and polarity orientation of the BMR. Although retuning the hyperparameters is not required, the model needs to be retrained whenever the initial conditions are modified. For standard linear advection–diffusion equations, such as the SFT model discussed here, the aforementioned limitation can be circumvented by using the Green's function approach (F. J. Leij et al. 2000; J. R. Bull 2016). For an initial condition of magnetic flux of strength unity at a given latitude and longitude, one can train the PINN to obtain the solution over the whole sphere and for $t > 0$. This procedure can be repeated over different latitudes and longitudes until a solution is obtained for each grid point. This forms a basis set, which can be used to express the general solution for any arbitrary initial condition. Additionally, this method can be applied to ensemble modeling to constrain the empirical parameters. Although this approach has been tested in a few cases, utilizing PINNs for this purpose is beyond the scope of this paper and will be addressed in future works.

We also tested multiple BMRs in the initial condition as a superposed input. The evolved magnetic field was found consistent with the sum of the individually evolved solutions, demonstrating the additivity property of the SFT equation. The versatility and robustness of the PINN method to different initial conditions emphasize its effectiveness and adaptability across a range of starting scenarios. Furthermore, we tested PINNs on observed magnetograms to reproduce the flux movement on the solar surface. This case study showcases the applicability of this novel approach to forecast magnetogram data that can be effectively used as an input to physics-based models like SWASTi (P. Mayank et al. 2022, 2023) and provides them with forecasting capabilities. In this work, we used initial conditions based on a functional form to simplify the analysis and emphasize key aspects of our model. However, incorporating real magnetogram data would offer a better understanding of the physical phenomena. We intend to integrate this into the model in our future projects.

The computational time taken for various methods for simulating the 1D case is as follows: (1) PINNs: 33 minutes on an NVIDIA GeForce RTX 3070 GPU for training and 0.6 s for generating solutions with the trained model, (2) RK-IMEX: 238 minutes on a CPU, and (3) Explicit: 152 minutes on a CPU.

In summary, we find that the PINN approach provides an excellent alternative to solving the SFT equation both in 1D and 2D. This pilot study compares PINNs with novel numerical schemes and showcases the tremendous potential of PINNs in providing accurate mesh-independent solutions. Considering its potential applicability in several areas of heliophysics, further studies are encouraged to develop more robust training methodologies that can adapt to changes in input data without requiring complete retraining.

### Acknowledgments

J.J.A. would like to express gratitude for the financial support received through the Prime Minister's Research Fellowship. B.V. and J.J.A. acknowledge the support received from the ISRO RESPOND grant No. ISRO/RES/2/436/21-22. S.K. acknowledges the support from STFC through grant ST/X001067/1. U.V. gratefully acknowledges support by NASA contracts NNG09FA40C (IRIS) and 80GSFC21C0011 (MUSE).

*Software:* DeepXDE (L. Lu et al. 2021), Matplotlib (J. D. Hunter 2007), and Numpy (C. R. Harris et al. 2020).

## Appendix
## Hyperparameter Optimization Using GP-based Bayesian Optimization

GP is a collection of random variables such that every finite subset has a multivariate normal distribution. GP-based Bayesian optimization is a powerful technique for hyperparameter optimization that strikes a balance between exhaustive grid search and random search. This method uses GP regression to guide the search for optimal hyperparameters by learning from previously evaluated configurations. There are two critical components (GP and acquisition function) of this method as discussed below. We provide a brief overview here, comprehensive descriptions and analytical expressions can be referred from P. I. Frazier (2018), T. Yu & H. Zhu (2020).

1. We model the prior distribution of the loss function ($\xi$) using a GP, with a particular mean and covariance function (Kernel function). We have used the Mátern kernel in this work. After evaluating $\xi$ for different hyperparameter sets, the posterior distribution is calculated using Bayes' rule. This posterior is the conditional distribution of the loss function $\xi$ given the already evaluated $\xi(\Theta)$. The posterior distribution gets updated with more evaluations, guided by the acquisition function.

2. The acquisition function guides the selection of the next set of hyperparameters where the loss ($\xi$) is to be evaluated. We have used the negative expected improvement (EI) function, defined as:

$$-\mathrm{EI}_n(\Theta) = -E_n[\max(\xi(\Theta) - \xi_n^+, 0)], \quad (A1)$$

where $\xi_n^+ = \min_{m \leqslant n} \xi(\Theta_m)$ represents the best point observed so far among $n$ evaluated hyperparameter sets. Operator $E_n$ calculates the expectation value under the posterior distribution of $\xi(\Theta)$. This function balances exploration (to parameter space that is not evaluated yet) and exploitation (to select the best parameter based on previous evaluations) by quantifying the expected gain from evaluating a new point.

**Algorithm 1.** Bayesian Optimization Using GP and EI

---

**Initialization:**
1. Randomly select an initial set of hyperparameters $\Theta_0$. Apply the hyperparameter set $\Theta_0$ to train the PINN network and calculate the loss $\xi(\Theta_0)$.
2. Build a GP prior distribution based on $(\Theta_0, \xi(\Theta_0))$

**Iteration:**
**while** the optimal configuration is not found and the resource limit is not reached **do**
   3. Calculate the acquisition function using the current posterior distribution and determine the next sampling point $\Theta^*$ corresponding to the minimum of the acquisition function.
   4. Apply the hyperparameter set $\Theta^*$ to train the PINN network and calculate the loss $\xi(\Theta^*)$.
   5. Update the posterior distribution (GP model) with the new data point $(\Theta^*, \xi(\Theta^*))$.
**EndWhile**
**Termination:**
6. Return the hyperparameter set corresponding to the least loss.

---

## ORCID iDs

Jithu J Athalathil ● https://orcid.org/0009-0002-9114-5880
Bhargav Vaidya ● https://orcid.org/0000-0001-5424-0059
Sayan Kundu ● https://orcid.org/0000-0003-3126-0588
Vishal Upendran ● https://orcid.org/0000-0002-9253-6093
Mark C. M. Cheung ● https://orcid.org/0000-0003-2110-9753